\documentclass[prd,preprintnumbers,superscriptaddress,twocolumn,aps,amsmath,amssymb,showpacs,nofootinbib]{revtex4-2}

\usepackage{graphicx,enumerate,xcolor,enumitem,xspace,braket,mathrsfs} 
\usepackage[colorlinks=true,citecolor=blue]{hyperref}
\usepackage{amsmath,amsfonts,amssymb}
\DeclareMathAlphabet{\mathbbold}{U}{bbold}{m}{n}

\usepackage{comment}
\usepackage{nicefrac}
\usepackage{multirow}
\usepackage{makecell}
\usepackage{mathtools}
\usepackage{hyperref}
\usepackage{url}
\usepackage{physics}   
\usepackage{dcolumn}
\usepackage{subfigure}
\usepackage{xspace}
\newcolumntype{e}[1]{D{+}{\,\pm\,}{#1}}
\newcolumntype{f}[1]{D{+}{~}{#1}}

\newcommand{\nn}{\nonumber\\}
\newcommand{\fb}{\operatorname{fb}}
\newcommand{\sh}{\mathrm{sh}}
\newcommand{\ch}{\mathrm{ch}}
\newcommand{\si}{\mathrm{s}}
\newcommand{\co}{\mathrm{c}}
\newcommand{\rhoM}{\rho_{\vphantom{\dot{b}}\hspace{-0.1em}\scalebox{0.5}{$M$}}}
\newcommand{\rhoMM}{\rho_{\vphantom{\dot{b}}\hspace{-0.1em}\scalebox{0.5}{$MM$}}}
\newcommand{\RMM}{\mathcal{R}_{\vphantom{\dot{b}}\hspace{-0.1em}\scalebox{0.5}{$MM$}}}

\newcommand{\mm}{$M \bar M$\xspace}
\newcommand{\kk}{$K^0 \bar K^0$\xspace}
\newcommand{\bb}{$B^0_s \bar B^0_s$\xspace}
\newcommand{\fcp}{f_{\eta_{CP}}}

\usepackage[normalem]{ulem}

\definecolor{nicegreen}{rgb}{0.,0.5,0.}

\newcommand{\customsection}[1]{\vspace{10pt}\noindent \textbf{#1}\vspace{5pt}}
\newcommand{\customsubsection}[1]{\vspace{5pt} \textit{#1}}

\begin{document}

\title{Quantum Tomography in Neutral Meson and Antimeson Systems}
\author{Kun Cheng}
\email{kun.cheng@pitt.edu}
\affiliation{PITT PACC, Department of Physics and Astronomy,\\ University of Pittsburgh, 3941 O’Hara St., Pittsburgh, PA 15260, USA}

\author{Tao Han}
\email{than@pitt.edu}
\affiliation{PITT PACC, Department of Physics and Astronomy,\\ University of Pittsburgh, 3941 O’Hara St., Pittsburgh, PA 15260, USA}

\author{Matthew Low}
\email{mal431@pitt.edu}
\affiliation{PITT PACC, Department of Physics and Astronomy,\\ University of Pittsburgh, 3941 O’Hara St., Pittsburgh, PA 15260, USA}

\author{Tong Arthur Wu}
\email{tow39@pitt.edu}
\affiliation{PITT PACC, Department of Physics and Astronomy,\\ University of Pittsburgh, 3941 O’Hara St., Pittsburgh, PA 15260, USA}

\date{\today}

\preprint{PITT-PACC-2503}

\begin{abstract}
The flavor space of particles produced in collider environments contains informative quantum correlations. We present a systematic approach for constructing the complete flavor density matrix for a meson and antimeson system (\mm) in the Bloch vector space at a given time $t$, which can be at or after production.  We point out that the $B_s^0$ and $K^0$ systems are superior to the $B^0_d$ and $D^0$ systems for quantum tomography because of their flavor oscillation and decay properties.  Performing quantum tomography for the \mm system can facilitate the study of production mechanisms, decoherence phenomena, quantum information variables, and potential new sources of CP violation. 
\end{abstract}

\maketitle

\customsection{Introduction}

Ever since the original quantum mechanical formulation of \kk mixing~\cite{Gell-Mann:1955ipe}, systems of a neutral meson $M$ and its antimeson $\bar{M}$ (\mm) have demonstrated many aspects of fundamental physics, such as the superposition of quantum states and the uncertainty principle.  These systems also probe physics at an energy scale much above the neutral meson masses through quantum loop effects. They have provided us with enriched laboratories that advance our understanding of quark masses, their flavor mixing, CP violation, as well as constraining new physics beyond the Standard Model (SM)~\cite{ParticleDataGroup:2024cfk}.

Due to the mixing between a meson and an antimeson, a meson state may be intrinsically treated as a qubit in flavor space.  Pairs of oscillating mesons correspond to two-qubit systems that have been measured for various mesons~\cite{CPLEAR:1997jgn,KLOE:2006iuj,KLOE-2:2021ila,Go:2003tx,Belle:2007ocp}. Attempts to observe entanglement and Bell nonlocality have been made, including in the \kk system from resonant $\phi$ decay~\cite{Six:1982kh,Selleri:1983nh,Eberhard:1992jd,DiDomenico:1995ky,Uchiyama:1996va,Selleri:1997vj,Foadi:1999sg,Foadi:2000zz,Benatti:1997fr,Benatti:1999du,Bertlmann:1999np,Bertlmann:2001ea,Bertlmann:2001sk,Bertlmann:2002wv,Bertlmann:2004yg,Li:2006fy,Ding:2007mk,Bramon:1998nz,Ancochea:1998nx,Gisin:2000jn,Dalitz:2000an,Bramon:2001tb,Hiesmayr:2000rm}, and in the $B^0_d \bar{B}^0_d$ system from the decay of the $\Upsilon(4S)$ particle~\cite{Datta:1986ut,Pompili:1999tv,Go:2003tx,Belle:2007ocp,Ichikawa:2008eq,Chen:2024drt}; see also Ref.~\cite{Atwood:2009un,Banerjee:2014vga,Shi:2016bvo,Shi:2019mlf,Takubo:2021sdk,Fabbrichesi:2025rqa}.  In these cases, the initial state has the definite quantum numbers $J^{PC}=1^{--}$, which means that the flavor state of the meson pair is a maximally entangled state as dictated by symmetry.  The existing results, however, focus mostly on comparing quantum and classical predictions and have limited use for quantum information.

In this work, we show how to reconstruct the most general two-particle flavor quantum state at an arbitrary time after production. We focus on the semi-leptonic decay channels in both the \bb and \kk systems, using the total semi-leptonic decay rate and the flavor asymmetry to form a complementary set of measurements of the flavor state. Based on the intrinsic properties of flavor oscillations and decays, we identify the optimal systems as $B_s^0$ and $K^0$ for quantum tomography. Using the observables we derive, we show that it is possible to fully measure the quantum properties of the flavor state of the meson pair, thus enabling, for the first time, systematic quantum-state tomography in flavor space. This work is largely orthogonal and complementary to the recent work on spin quantum tomography~\cite{Afik:2020onf,Barr:2021zcp,Fabbrichesi:2021npl,Severi:2021cnj}.

When a heavy quark pair originates from open flavor production at colliders, via the strong or electromagnetic interactions, the corresponding heavy hadrons are initially in flavor eigenstates. In general, however, the flavor correlation of the pair of hadronic states cannot be determined from first principles because of the non-perturbative hadronization process.  Developing an approach for quantum tomography in this system may therefore provide an opportunity to study the underlying hadronization mechanisms \cite{vonKuk:2025kbv}.

The demonstration of flavor quantum tomography not only makes all quantum information measurements, such as concurrence, quantum discord, and quantum entropies available in the flavor sector, but also provides a crucial reference for theoretical predictions of the flavor state production from non-perturbative processes, as well as potential observations of new sources of CP violation.

\customsection{Flavor Quantum States and Time-Evolution}

When a meson is neutral under gauge symmetries, the meson state $\ket{M}$ mixes with the antimeson state $\ket{\bar M}$.  Examples of neutral mesons $M$ include $B_d^0$, $B_s^0$, $D^0$, and $K^0$.  
In the flavor space spanned by $\ket{M}$ and $\ket{\bar M}$, a single meson can be analyzed as a two-level quantum system~\cite{Gell-Mann:1955ipe}, known as a qubit, and described with a density matrix $\rhoM$.  This matrix can be parameterized as
\begin{equation}
\label{eq:rhosingle}
\rhoM = \frac{\mathbbold{1}_2 + b_i \sigma_i}{2},
\end{equation}
where $\mathbbold{1}_2$ is the two-dimensional identity matrix and $\sigma_i$ ($i=x,y,z$) are the Pauli matrices, with repeated indices summed over.  The real vector $b_i$, known as the Bloch vector, specifies a quantum state in the Bloch vector space~\cite{Bloch:1946zza,Feynman}. 

With a common mass $m$ and a common decay width $\Gamma$ for the meson and antimeson, the non-Hermitian effective Hamiltonian is
\begin{equation}
\label{eq:massmatrix_nondiag}
\mathcal{H}=
\mathbf{M}-i \boldsymbol{\Gamma}/2=
\begin{pmatrix}
  m - i\frac{\Gamma}{2} & H_{12} \\
  H_{21} &  m - i\frac{\Gamma}{2} 
\end{pmatrix},
\end{equation}
where $H_{12}$ and $H_{21}$ are complex parameters that encode meson-antimeson mixing.  Diagonalizing Eq.~\eqref{eq:massmatrix_nondiag} yields the mass eigenstates $\ket{M_1}$ and $\ket{M_2}$, along with their masses and decay widths:
\begin{align}
    \ket{M_1} &= p \ket{M} + q\ket{\bar M},  \qquad \text{with }(m_1,\;\Gamma_1),   \\
    \ket{M_2} &= p \ket{M} - q \ket{\bar M}, \qquad \text{with } (m_2,\;\Gamma_2),
\end{align}
where $|q|^2 + |p|^2 = 1$ and $p/q = \sqrt{H_{12}/H_{21}}$.  In the absence of CP violation $p=q=1/\sqrt{2}$.  We adopt the convention $\text{CP} \ket{M}=\ket{\bar{M}}$, implying that $\ket{M_1}$ is a CP-even eigenstate and $\ket{M_2}$ is CP-odd.  In the Bloch vector space, the flavor eigenstates $\ket{M}$ and $\ket{\bar M}$ are described by $\vec{b}=(0,0,\pm1)$ while the mass eigenstates $\ket{M_1}$ and $\ket{M_2}$ are described by $\vec{b}=(\pm1,0,0)$. When CP is not conserved the eigenstates of CP and of mass no longer coincide.

Since the meson or antimeson can decay, as evidenced by their non-zero widths, this is an open quantum system that can be effectively described by non-unitary time evolution $U(t)$~\cite{Caban:2006ij,Roccati:2022yjo,Karamitros:2022oew,Alok:2024amd},
\begin{widetext}
\begin{equation}\label{eq:EvolutionMatrixU}
    U(t)= \begin{pmatrix}
        \frac{1}{2}(e^{-\Gamma_1 t/2-i m_1 t}+e^{-\Gamma_2 t/2-i m_2 t}) & \frac{q}{2p}(e^{-\Gamma_1 t/2-i m_1 t} - e^{-\Gamma_2 t/2-i m_2 t}) \\
        \frac{p}{2q}(e^{-\Gamma_1 t/2-i m_1 t} - e^{-\Gamma_2 t/2-i m_2 t}) & \frac{1}{2}(e^{-\Gamma_1 t/2-i m_1 t}+e^{-\Gamma_2 t/2-i m_2 t}) 
    \end{pmatrix},
\end{equation}
\end{widetext}
where $U(t)$ is expressed in the flavor basis.  The fraction of remaining mesons at time $t$ is
\begin{equation}
\frac{N(t)}{N_0}=\tr (U(t) \rhoM U(t)^\dagger),
\end{equation}
where $N_0$ is the number of mesons at $t=0$ and $N(t)$ is the number of mesons left at time $t$. The explicit form of $N(t)$ is given in Appendix~\ref{app:evo1}.  One can choose any starting point as $t=0$ such as the time of production.  
 
The time evolution of the quantum density matrix, in the Schr\"odinger picture, is
\begin{equation}
\label{eq:rhoMt}
\rhoM(t) = \frac{U(t) \rhoM U(t)^\dagger}{\tr (U(t) \rhoM U(t)^\dagger)}. 
\end{equation}
The normalization of $\rhoM(t)$ leads to unitary evolution.

\customsection{Oscillation and Decay in the Bloch Vector Space}

It is intuitive to analyze the state evolution in the Bloch vector space~\cite{Bloch:1946zza,Feynman}, because different components of the quantum density matrix $\rhoM$ are probed by the various decays of the neutral meson, as we formulate explicitly below.  The time evolution of the Bloch vector is~\cite{Kerbikov:2017spv,Karamitros:2022oew,Karamitros:2025azy} 
\begin{equation}
\label{eq:OperatorPrecession}
\frac{d}{dt}  \vec b(t) = - 2\vec E \times \vec b(t) + \vec{\Gamma}-[\vec{\Gamma}\cdot \vec{b}(t)] \; \vec{b}(t),
\end{equation}
where the vectors $\vec{E}$ and $\vec\Gamma$ are defined from the decompositions of the Hermitian matrices $\mathbf{M}$ and $\mathbf{\Gamma}$ in Eq.~\eqref{eq:massmatrix_nondiag} as
\begin{equation}
\mathbf{M}=E_0\mathbbold{1}_2+\vec{E}\cdot \vec{\sigma},
\qquad\qquad
\mathbf{\Gamma}=\Gamma_0\mathbbold{1}_2+\vec{\Gamma}\cdot \vec{\sigma}.
\end{equation}
In this paper, we work in the CP-conserving limit with
\begin{equation}
H_{21}=H_{12} 
= \frac{\Delta m }{2}+i \frac{\Delta \Gamma}{4},
\end{equation}
where $\Delta m\equiv m_2-m_1$  and $\Delta\Gamma=\Gamma_1-\Gamma_2$.
Therefore, both vectors $\vec{E}=(\Delta m/2,0,0)$ and $\vec{\Gamma}=(-\Delta \Gamma/2,0,0)$ point along the $\vec x$ direction. 

\customsubsection{Collapse to Flavor Eigenstates} $-$ Consider a multi-particle final state $\ket{f}$ with ${\rm CP}\ket{f}= \ket{\bar f}$ and $\ket{f}\neq \ket{\bar f}$. In the CP-conserving limit, the meson $M$ only decays to $f$ while the antimeson $\bar M$ only decays to $\bar f$. The semi-leptonic decays of neutral mesons are examples of such decays. The observation of such decays identifies the flavor of the neutral meson as $\ket{M}$ or $\ket{\bar{M}}$.

The probabilities of collapsing the meson state to flavor eigenstates are
\begin{equation}
\bra{M}\rhoM(t)\ket{M}=\frac{1+b_z(t)}{2},
\ \ 
\bra{\bar M}\rhoM(t)\ket{\bar M}=\frac{1-b_z(t)}{2}.
\nonumber
\end{equation}
Let $N_{f/\bar f}(t)$ be the number of $f/\bar f$ final states produced in decays from $t=0$ until time $t$.  The rate of change is
\begin{equation}
\frac{d}{dt} N_{f/\bar f}(t)
= N(t)  \frac{1\pm b_z(t)}{2} \Gamma_{M\to f},
\end{equation}
where $\Gamma_{M\to f}$ is the partial width of $M$ decaying to $f$ and $\Gamma_{M\to f}=\Gamma_{\bar M\to \bar f}$.

In practice, we propose to measure the decay asymmetry $N_f(t)-N_{\bar f}(t)$, which directly corresponds to the expectation value of $\sigma_z$ with $\Tr(\sigma_z\!\cdot\! \rhoM(t))=b_z(t)$,
\begin{align}
\frac{d(N_{f}(t)-N_{\bar f}(t))}{dt}&= N(t)  b_z(t) \Gamma_{M\to f},  \nonumber\\
    &=N_0\, e^{-\Gamma t} (b_z \co_t-b_y \si_t) \Gamma_{M\to f} .
    \label{eq:trsigmaz}
\end{align}
where $\co_t=\cos(\Delta m t)$,  $\si_t=\sin(\Delta m t)$, and $b_{i}=b_{i}(0)$ are the components of the Bloch vector at $t=0$.
The decay asymmetry probes both $b_y$ and $b_z$ as they evolve into each other according to Eq.~\eqref{eq:OperatorPrecession}, which describes precession around the $x$ direction in the limit of $\Delta m \gg \Gamma$. 
Therefore, for different events within the ensemble, decays occur at different times, allowing us to measure different directions in the $y-z$ plane of the Bloch vector space.

In the Heisenberg picture, the Bloch vector $\vec b(0)$ does not rotate.  Instead, the asymmetry operator $\sigma_z$ oscillates with $\sigma_y$.  Measuring decays at different times probes the operator at different times and maps out the $y$ and $z$ components of the Bloch vector at $t=0$.

\customsubsection{Collapse to CP eigenstates} $-$ The neutral mesons can also decay to CP eigenstates with $\ket{f_{\eta_{CP}}}= \text{CP}\ket{f_{\eta_{CP}}}=\eta_{CP} \ket{f_{\eta_{CP}}}$, where $\eta_{CP}=\pm 1$ is the CP eigenvalue of the final state $\ket{f_{\eta_{CP}}}$.  The decays of $B_s^0 \to J/\psi\eta$ and $K^0\to \pi^+\pi^-$ are such examples.  Instead of collapsing the meson state to $\ket{M}$ or $\ket{\bar M}$, the decay to $\ket{\fcp}$ collapses the meson state to the CP eigenstates (or mass eigenstates) $\ket{M_1}$ or $\ket{M_2}$, whose Bloch vectors are along the $x$ 
direction.  Such decays allow the reconstruction of $b_x$.

Due to the uneven decay rates of the CP-odd and CP-even states, 
there is no simple asymmetric observable directly corresponding to $\langle\sigma_x\rangle$ as in Eq.~\eqref{eq:trsigmaz}. However, the decay can still be viewed as a projection to $\ket{M_1}$ and $\ket{M_2}$  
with\footnote{We assume CP-conserving $\langle{M|f_+}\rangle=\langle{\bar{M}|f_+}\rangle$ for illustration while our result Eq.~\eqref{eq:Nf+Nf} does not rely on this.}
\begin{equation}
\bra{M_1}\rhoM(t)\ket{M_1}=\frac{1+b_x(t)}{2},~
\bra{M_2}\rhoM(t)\ket{M_2}=\frac{1-b_x(t)}{2}. \nonumber
\end{equation}
Therefore, the $x$ component of the Bloch vector can be reconstructed from the decay to CP eigenstates.  Furthermore, the decay to CP eigenstates results in a decay width difference $\Delta\Gamma=\Gamma_1-\Gamma_2$ and lifetime difference, leading to the $b_x$ dependence in the decay rate to the total flavor eigenstates, both $f$ and $\bar f$, as a function of time 
\begin{align}
    \frac{d(N_{f}(t)+N_{\bar f}(t))}{dt} &= N(t) \Gamma_{M\to f} \nn
    &=N_0\, e^{-\Gamma t}  (\ch_t-b_x\sh_t)  \Gamma_{M\to f} , 
    \label{eq:Nf+Nf}
\end{align}
where $\ch_t=\cosh(\Delta\Gamma t/2)$ and  $\sh_t=\sinh(\Delta\Gamma t/2)$.  Thus, one can fully reconstruct $(b_x, b_y, b_z)$ by only measuring the $M\to f/\bar f$ decay channels and comparing these observables with Eqs.~\eqref{eq:trsigmaz} and~\eqref{eq:Nf+Nf}.

In summary, the components of the Bloch vector in the $y-z$ plane are probed by the flavor asymmetry $N_f~-~N_{\bar f}$ through oscillations in Eq.~\eqref{eq:trsigmaz}, and the Bloch vector in the $x$ direction is measured from $N_f+N_{\bar f}$ via total decay rates in Eq.~\eqref{eq:Nf+Nf}. By fitting the distributions of $N_f \pm N_{\bar f}$ with respect to decay time, the complete flavor density matrix for a single meson is reconstructed.

\customsubsection{Generalization to Meson Pairs} $-$ The quantum state of a two-meson system, with the two subsystems denoted as $\mathcal{A}$ and $\mathcal{B}$, respectively, can be expressed by the density matrix 
\begin{equation}
\label{eq:rhomesonPair}
\rhoMM = \frac{\mathbbold{1}_4 + b_i^{\mathcal{A}} \sigma_i\otimes \mathbbold{1}_2 + b_i^{\mathcal{B}} \mathbbold{1}_2 \otimes \sigma_i + C_{ij} \sigma_i\otimes \sigma_j}{4}, 
\end{equation}
\\[-19mm]
\begin{equation}
\phantom{a} \notag
\end{equation}
where $b^{\mathcal{A}}_i$ and $b^{\mathcal{B}}_i$ describe the flavor state of mesons $\mathcal{A}$ and $\mathcal{B}$, respectively, and $C_{ij}$ is the flavor correlation matrix between the two mesons.  Repeated indices are summed over.  Collectively, these are also called the Fano coefficients~\cite{Fano:1983zz}.

Like the single meson case, the Fano coefficients of the two-meson system $b^{\mathcal{A}}_i$,  $b^{\mathcal{B}}_i$, and $C_{ij}$ are obtained by measuring the flavor asymmetry and the total decay rate to a flavor eigenstate, for each meson.  This leads us to define the following four observables
\begin{align}
N_{\rm tot} &= N_{f f} +  N_{\bar f f} + N_{f \bar f} + N_{\bar f \bar f}, \\
A_{ff}&= N_{f f} -  N_{\bar f f} - N_{f \bar f} + N_{\bar f \bar f}, \\
A^{\mathcal{A}}_{f}&=N_{f f}  - N_{\bar f f} +  N_{f \bar f} - N_{\bar f \bar f},  \\ 
A^{\mathcal{B}}_{f}&=N_{f f} + N_{\bar f f} -  N_{f \bar f} - N_{\bar f \bar f},
\end{align}
where $N_{f_1 f_2}$ is the number of events in which  meson $\mathcal{A}$ decays to $f_1$ and meson $\mathcal{B}$ decays to $f_2$.  All quantities above depend on the times $t_1$ and $t_2$.  

$N_{\rm tot}$ is the total number of events from semi-leptonic decays, $A_{ff}$ is the asymmetry between same-flavor and opposite flavor decays, and $A_{f}^{\mathcal{A}/\mathcal{B}}$ is the decay flavor asymmetry of one meson under the condition that the other meson also decays to a flavor eigenstate.  The distributions of the four observables have the following dependence on the flavor density matrix of the meson pair (see Appendix~\ref{app:evo2} for details), \vspace{-10mm}
\begin{widetext}
\begin{align}
    \frac{\dd^2 N_{\rm tot}}{\dd t_1 \dd t_2} & = N_0 \Gamma_{B_0\to f}^2\, e^{-\Gamma(t_1+t_2)} \big(\ch_{t_1}\ch_{t_2}- \ch_{t_1}\sh_{t_2}b^{\mathcal{A}}_x- \sh_{t_1}\ch_{t_2}b^{\mathcal{B}}_x + C_{xx} \sh_{t_1} \sh_{t_2} \big) + \mathcal{O}(\epsilon), \label{eq:Ntot} \\
    \frac{\dd^2 A_{ff}}{\dd t_1 \dd t_2} & = N_0 \Gamma_{B_0\to f}^2 \, e^{-\Gamma(t_1+t_2)} \big(\si_{t_1}\si_{t_2}C_{yy} -\co_{t_1}\si_{t_2}C_{zy}- \si_{t_1}\co_{t_2}C_{yz}+\co_{t_1}\co_{t_2}C_{zz} \big) + \mathcal{O}(\epsilon), \label{eq:A} \\
    \frac{\dd^2 A^{\mathcal{A}}_{f}}{\dd t_1 \dd t_2} & = N_0 \Gamma_{B_0\to f}^2\, e^{-\Gamma(t_1+t_2)} \big( \ch_{t_2}(\co_{t_1}b^{\mathcal{A}}_z-\si_{t_1}b^{\mathcal{A}}_y) -\sh_{t_2}(\co_{t_1}C_{zx}-\si_{t_1}C_{yx}) \big)  + \mathcal{O}(\epsilon), \label{eq:AA} \\ 
    \frac{\dd^2 A^{\mathcal{B}}_{f}}{\dd t_1 \dd t_2} & = N_0 \Gamma_{B_0\to f}^2\, e^{-\Gamma(t_1+t_2)} \big(\ch_{t_1}(\co_{t_2}b^{\mathcal{B}}_z -\si_{t_2}b^{\mathcal{B}}_y ) -\sh_{t_1}(\co_{t_2}C_{xz} - \si_{t_2}C_{xy}) \big) + \mathcal{O}(\epsilon),  \label{eq:AB}
\end{align}
\end{widetext}
where $N_0$ is the total number of $M\bar{M}$ events, $t_1$ and $t_2$ are the decay times of the first and second mesons, respectively.  Here, $\epsilon$ is the CP violation parameter in the \mm  mixing defined by ${p}/{q}=({1+\epsilon})/({1-\epsilon})$. For $B_s^0$, $\epsilon\sim\mathcal{O}(10^{-5})$, and for $K^0$, $\epsilon\sim \mathcal{O}(10^{-3})$. This validates our analysis as a good approximation in the CP-conserving limit. Another source of CP violation, the difference between the decay amplitudes of $M$ and $\bar M$ when decaying to CP eigenstates~\cite{Pich:1993et,Buchalla:1997jn}, is irrelevant to the four observables in the semi-leptonic decay channel above.

By measuring the four distributions, we reconstruct all the Fano coefficients $\vec{b}^{\mathcal{A}}$, $\vec{b}^{\mathcal{B}}$, and $C_{ij}$. Deviations from the predictions made by the density matrix parametrization may indicate the presence of decoherence phenomena~\cite{KLOE:2006iuj,KLOE-2:2021ila,Alok:2024amd}. 

\customsection{Simulations and Experimental Feasibility}

From an observational perspective, an \mm system needs to have a sizable value of $\Delta m/\Gamma$ and $\Delta \Gamma/\Gamma$, ideally of order unity or larger.  This corresponds to rapid oscillations and an appropriate decay lifetime difference.

As shown in Table~\ref{tab:parameter}, the $B_s^0$ system is such a system with $(\Delta m/\Gamma,\ \Delta \Gamma/\Gamma)_{B_s^0}=(27,\ 0.14)$. 
The $B_d^0$ system, with $(\Delta m/\Gamma,
\ \Delta \Gamma/\Gamma)_{B_d^0}=(0.77,\  0.004)$, does have a suitable oscillation time, but the decay lifetime difference is too small to yield a usable measurement of $b_x$. However, in this system, the $(2\times 2)_{yz}$ subset of the correlation matrix can be measured, which is sufficient to measure Bell nonlocality~\cite{Go:2003tx,Belle:2007ocp}, but full tomography cannot be performed.  The $D^0$ system has small values of both $\Delta m/\Gamma$ and $\Delta \Gamma/\Gamma$ which makes quantum tomography much more challenging.  For the $K^0$ system both quantities are sizable, $(\Delta m/\Gamma,\ \Delta \Gamma/\Gamma)_{K^0}=(0.95,\ 1.99)$.  We therefore adopt the $B_s^0$ and $K^0$ systems to quantitatively demonstrate the procedure of quantum tomography. 

For both the $B_s^0$ and $K^0$ systems, we choose two examples of density matrices for concreteness.  The first is $\rho_{\rm Bell}$, parametrized as 
\begin{equation}
    b^\mathcal{A}_i=b^\mathcal{B}_i=0 \quad {\rm  and}\quad C_{ij}=-\delta_{ij},
    \label{eq:init}
\end{equation}
which corresponds to the CP-odd flavor state $(\ket{M\bar M}-\ket{\bar M M})/\sqrt{2}$. The second is the Werner state $\rho_\kappa$~\cite{PhysRevA.40.4277} given by
\begin{equation}
\rho_\kappa = (1-\kappa)\rho_{\rm Bell} + \kappa \frac{\mathbbold{1}_4}{4},
\label{eq:decoh}
\end{equation} 
where $\mathbbold{1}_4/4$ is a maximally mixed state and we use $\kappa=0.2$.

The samples are distributed in the two-dimensional space $(t_1,t_2)$ where $t_1$ and $t_2$ are the decay times of the first and second mesons, respectively.  For each of the two states $\rho_{\rm Bell}$ and $\rho_\kappa$, we consider the binned distribution of $N_{\rm tot}$ from $10^7$ semi-leptonically-decaying events.  From this we extract $N_0$, $b^{\mathcal{A}}_x$, $b^{\mathcal{B}}_x$, and $C_{xx}$ with Eq.~\eqref{eq:Ntot}.  The extracted value of $N_0$ is then used when fitting to $A_{ff}$, $A^{\mathcal{A}}_{f}$, and $A^{\mathcal{B}}_{f}$, in Eqs.~\eqref{eq:A}$-$\eqref{eq:AB}, to measure the coefficients $b_i^{\mathcal{A}}$, $b_i^{\mathcal{B}}$, and $C_{ij}$.   The central value and statistical uncertainty of fitted results are estimated from the mean and the standard deviation of 40 pseudo-experiments. 

\begin{figure} 
  \centering
 \includegraphics[width=0.9\linewidth]{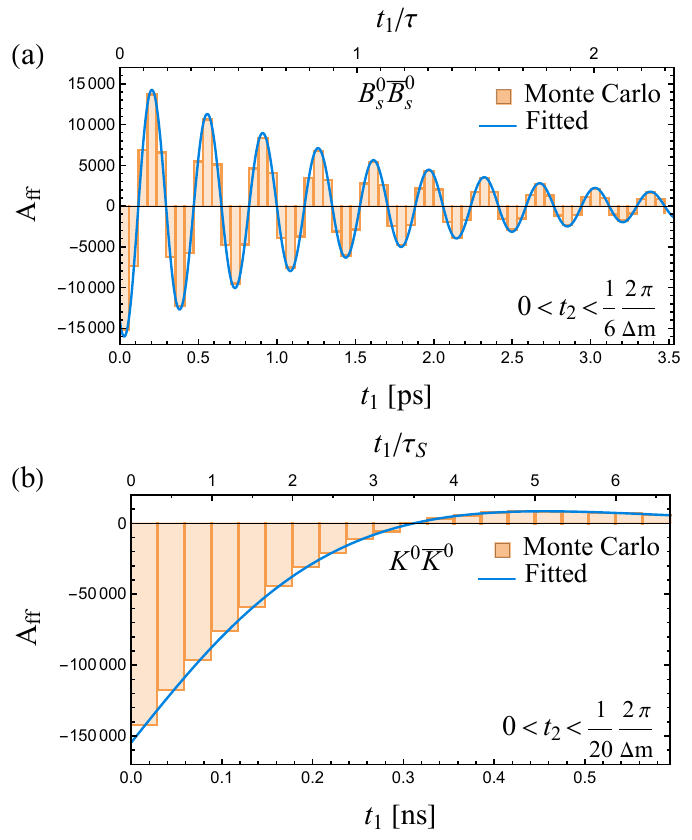}
  \caption{Monte Carlo results and fitted distribution of $A_{ff}(t_1,t_2)$ for $B^0_s \bar{B}^0_s$ and $K^0 \bar{K}^0$, in the direction of $t_1$ with $t_2$ interval specified in the figure.  The labels of the upper axes are the time in units of the decay 
  lifetime of $B^0_s$ and $K^0_S$.}
\label{fig:fit}
\end{figure}

The $B_s^0$ meson pair events are generated with $0<t_1,t_2<3.5\;\text{ps}$, within the first 10 periods of oscillation and about twice the mean lifetime. The oscillatory time evolution versus $t_1$ is shown in Fig.~\ref{fig:fit}(a), integrated over a range of $t_2$ for the $\rho_{\rm Bell}$ initial state.  We adopt a similar bin width as Ref.~\cite{LHCb:2021moh} with each period divided into 6 bins, leading to $60\times 60$ bins in the $t_1-t_2$ plane. We find that the density matrix can be reconstructed to high precision in our formalism, as shown in Table~\ref{tab:fit}.  The measurement precision of the $C_{xx}$ entry of the flavor correlation matrix is much lower than the other coefficients, because the $B_s^0$ mesons have relatively small $\Delta\Gamma/\Gamma$ and the sensitivity of $C_{xx}$ is suppressed by two factors of $\sinh(\Delta\Gamma t/2)$.

The analysis for $K^0$ meson pairs is similar.  We generate events with $0<t_1,t_2<0.59\;\text{ns}$, within the first half-period of the damped oscillation and 6.7 times the lifetime of $K_S$, as shown in Fig.~\ref{fig:fit}(b).  We divide the half-period into $20$ bins, similar to Ref.~\cite{CPLEAR:1999bft}, which yields a total $20\times20$ bins.  Although $K^0$ and $\bar K^0$ cannot completely evolve into each other before the quick $K_S$ decay, the oscillation of about a quarter-period is still sufficient to extract $b_{y}$ and $b_{z}$. With nearly maximal $\Delta\Gamma/\Gamma$, the sensitivity of the $C_{xx}$ coefficient is not suppressed in the $K^0$ system and can be reconstructed equally well as the other coefficients. 
Collecting events with decays extending to several times the $K_S$ lifetime is enough to perform quantum tomography, although studying the $K_L$ events at later times is also fruitful in exploring quantum phenomena~\cite{KLOE:2006iuj,KLOE-2:2021ila}.

\begin{table}[tb]
\renewcommand{\arraystretch}{1.4}
    \centering
    \begin{tabular}{|c|f{8.11}|f{8.11}|c|}
    \hline
         & \multicolumn{1}{c|}{$B^0_s\bar{B}^0_s$ fitted} & \multicolumn{1}{c|}{$K^0\bar{K}^0$ fitted} & \multirow{2}{*}{Obs.} \\[-1mm]
         & \multicolumn{1}{c|}{$\rho_{\rm Bell}$ ($\rho_\kappa$)} & \multicolumn{1}{c|}{$\rho_{\rm Bell}$ ($\rho_\kappa$)} & \\
    \hline
    \hline
        $b^{\mathcal{A}}_x$ & \scalebox{0.85}{$0{\pm 0.020}$}+\scalebox{0.85}{$(0{\pm 0.017})$} & \scalebox{0.85}{$0{\pm 0.0015}$} +  \scalebox{0.85}{$(0{\pm 0.0018})$} & \multirow{2}{*}{$N_{\rm tot}$} \\
    \cline{1-3}
        $b^{\mathcal{B}}_x$ & \scalebox{0.85}{$0{\pm 0.019}$} +  \scalebox{0.85}{$(0{\pm 0.016})$} & \scalebox{0.85}{$0{\pm 0.0017}$} +  \scalebox{0.85}{$(0{\pm 0.0018})$} &  \\
    \hline
        $b_y^{\mathcal{A}}$ & \scalebox{0.85}{$0{\pm 0.0009}$} +  \scalebox{0.85}{$(0{\pm 0.0010})$} & \scalebox{0.85}{$0{\pm 0.0019}$} +  \scalebox{0.85}{$(0{\pm 0.0021})$} & \multirow{2}{*}{$A^{\mathcal{A}}_{f}$} \\
    \cline{1-3}
        $b_z^{\mathcal{A}}$ & \scalebox{0.85}{$0{\pm 0.0010}$} +  \scalebox{0.85}{$(0{\pm 0.0009})$} & \scalebox{0.85}{$0{\pm 0.0014}$} +  \scalebox{0.85}{$(0{\pm 0.0016})$} & \\
    \hline
        $b_y^{\mathcal{B}}$ & \scalebox{0.85}{$0{\pm 0.0010}$} +  \scalebox{0.85}{$(0{\pm 0.0008})$} & \scalebox{0.85}{$0{\pm 0.0020}$} +  \scalebox{0.85}{$(0{\pm 0.0023})$} & \multirow{2}{*}{$A^{\mathcal{B}}_{f}$} \\
    \cline{1-3}
        $b_z^{\mathcal{B}}$ & \scalebox{0.85}{$0{\pm 0.0009}$} +  \scalebox{0.85}{$(0{\pm 0.0008})$} & \scalebox{0.85}{$0{\pm 0.0015}$} +  \scalebox{0.85}{$(0{\pm 0.0016})$} & \\
    \hline
    \hline
        $C_{xx}$ & \scalebox{0.85}{$-1{\pm 0.36}$} +  \scalebox{0.85}{$(-0.8{\pm 0.29})$} & \scalebox{0.85}{$-1{\pm 0.0031}$} + \scalebox{0.85}{$(-0.8{\pm 0.0032})$} & $N_{\rm tot}$  \\
    \hline
        $C_{yx}$ & \scalebox{0.85}{$0{\pm 0.016}$} +  \scalebox{0.85}{$(0{\pm 0.018})$} & \scalebox{0.85}{$0{\pm 0.0023}$} +  \scalebox{0.85}{$(0{\pm 0.0026})$} & \multirow{2}{*}{$A^{\mathcal{A}}_{f}$} \\
    \cline{1-3}
        $C_{zx}$ & \scalebox{0.85}{$0{\pm 0.018}$} +  \scalebox{0.85}{$(0{\pm 0.017})$} & \scalebox{0.85}{$0{\pm 0.0017}$} +  \scalebox{0.85}{$(0{\pm 0.0019})$} & \\
    \hline
        $C_{xy}$ & \scalebox{0.85}{$0{\pm 0.017}$} +  \scalebox{0.85}{$(0{\pm 0.015})$} & \scalebox{0.85}{$0{\pm 0.0025}$} +  \scalebox{0.85}{$(0{\pm 0.0029})$} & \multirow{2}{*}{$A^{\mathcal{B}}_{f}$} \\
    \cline{1-3}
        $C_{xz}$ & \scalebox{0.85}{$0{\pm 0.016}$} +  \scalebox{0.85}{$(0{\pm 0.016})$} & \scalebox{0.85}{$0{\pm 0.0018}$} +  \scalebox{0.85}{$(0{\pm 0.0020})$} & \\
    \hline
        $C_{yy}$ & \scalebox{0.85}{$-1{\pm 0.0011}$} +  \scalebox{0.85}{$(-0.8{\pm 0.0010})$} & \scalebox{0.85}{$-1{\pm 0.0027}$} +  \scalebox{0.85}{$(-0.8{\pm 0.0023})$} & \multirow{4}{*}{$A_{ff}$} \\
    \cline{1-3}
        $C_{yz}$ & \scalebox{0.85}{$0{\pm 0.0007}$} +  \scalebox{0.85}{$(0{\pm 0.0008})$} & \scalebox{0.85}{$0{\pm 0.0022}$} +  \scalebox{0.85}{$(0{\pm 0.0020})$} &  \\
    \cline{1-3}
        $C_{zy}$ & \scalebox{0.85}{$0{\pm 0.0008}$} +  \scalebox{0.85}{$(0{\pm 0.0008})$} & \scalebox{0.85}{$0{\pm 0.0021}$} +  \scalebox{0.85}{$(0{\pm 0.0017})$} &  \\
    \cline{1-3}
        $C_{zz}$ & \scalebox{0.85}{$-1\!\!\pm\!\!0.0011$} +  \scalebox{0.85}{$(-0.8\!\!\pm\!\!0.0009)$}  & \scalebox{0.85}{$ -1{\pm 0.0010}$} +  \scalebox{0.85}{$(-0.8{\pm 0.0011})$} & \\
    \hline
    \end{tabular}
    \caption{The central values and statistical uncertainties of the Fano coefficients in the $B_s^0\bar B_s^0$ system and in the $K^0\bar{K}^0$ system, using the initial states $\rho_{\rm Bell}$ and $\rho_\kappa$ with $\kappa=0.2$.  Each system uses $10^7$ events and 40 pseudo-experiments.  The statistical uncertainties scale as $1/\sqrt{N}$. }
    \label{tab:fit}
\end{table}

Results are shown in Table~\ref{tab:fit} with $10^7$ events.  In practice, approximately $7\times 10^6$ $B_s^0\bar{B}_s^0$ pairs have been produced at Belle~\cite{Glattauer:2015yag,Belle:2008ezn}, and around $3\times10^8$ are expected to be produced at Belle II~\cite{Oswald:2013tna}, mostly in pure Bell states from the decay of the $\Upsilon(5S)$. 

At LHCb, a pair of $B^0_s$ mesons can be produced from the hadronization of $b\bar{b}$, which has a cross section of $\sigma(b\bar{b}X)=0.56~\mathrm{mb}$~\cite{dEnterria:2016ids}, and a fraction of around $1.4\%$~\cite{LHCb:2019fns} goes to a pair of $B^0_s$ mesons. With the Run 2 luminosity of $5.7~\fb^{-1}$, there are $8\times10^{10}$ $B^0_s\bar{B}^0_s$'s produced. The fraction of both $B^0_s$'s decaying semi-leptonically is $\mathrm{Br}(B_s^0\to \ell^+ \nu_\ell X^-)^2=3.7\%$, which potentially yields as many events as our sample simulation. The reconstruction of a pair of $b$-hadrons has been discussed in Ref.~\cite{Afik:2024uif}, and we expect a similar efficiency. Due to the complicated nature of production and hadronization in hadronic collisions, it is unclear to what extent a $B^0$ meson pair would be flavor-entangled, which is very interesting for future work, both in theory and in experiment. 

$K^0$ is typically pair produced from a $\phi$ meson decay. Around $8\times10^9$ $K^0 \bar{K}^0$ pairs have been produced at KLOE and KLOE-2 \cite{DiDomenico:2020pxd}, and a fraction of $1.1\%$ decays in $0<t< \pi /\Delta m$, among which a fraction of $\mathrm{BR}(K^0_S \to \pi^\pm \ell^\mp \nu_\ell)\mathrm{BR}(K^0_L \to \pi^\pm \ell^\mp \nu_\ell)=8\times10^{-4}$ decays to leptons on both sides.

\customsection{Applications to Quantum Information}

Having performed quantum tomography for the \mm system, we are in a position to study any quantum information observable of the system.  We list some example quantities below and provide their definitions in Appendix~\ref{app:OQM}. 

Concurrence is a measure of the entanglement of a system which measures how much the two subsystems are not separable~\cite{Einstein:1935rr}.  The Bell variable is a normalized measurement of Bell's inequality where a positive value signifies that a local hidden variable model cannot imitate this result~\cite{Bell:1964kc}.

Quantum discord captures a broader class of quantum correlations than entanglement; it can be non-zero even for separable states~\cite{Zurek_2000,Ollivier:2001fdq}.  A non-zero value of discord indicates that the system is not invariant under measurements, which is one of the prime features of quantum systems. The steerability variable is a normalized measure of steerability where a positive value indicates that the state is steerable.  This is a weaker condition than Bell nonlocality that corresponds to whether the state admits a local hidden-state model, where one party’s measurements are classically explainable even if the other’s are not.

Conditional entropy quantifies the number of bits needed for subsystem $\mathcal{A}$ to reconstruct subsystem $\mathcal{B}$.  Classically, this is restricted to non-negative values, but quantum mechanically negative values are possible~\cite{Horodecki:2005vvo}.
The second stabilizer R\'enyi entropy (SSRE) is a measure of the magic of a quantum system which quantifies the computation advantage this system would have over a classical computer~\cite{Leone:2021rzd}.

These quantities are listed in Table~\ref{tab:quantumObservables} for $\rho_\kappa$ with $\kappa=0.2$.  All quantities can be measured to a high precision.  For quantities that require the full density matrix for their calculation, their precision is limited by the largest uncertainty in Table~\ref{tab:fit}.  Some quantities, however, only require a subset of the Fano coefficients, like Bell nonlocality, in which case it can often be measured to higher precision.

\begin{table}[tbh]
    \centering
    \begin{tabular}{|c|e{6.6}|e{6.6}|}
    \hline
        \rule{0pt}{2.4ex}  & \multicolumn{1}{c|}{$B^0_s\bar{B}^0_s$}  & \multicolumn{1}{c|}{$K^0\bar{K}^0$}  \\
    \hline
        Concurrence & 0.69 + 0.14 & ~0.702 + 0.0024  \\
    \hline
        Bell Variable & 0.1854 + 0.0014~ & 0.188  + 0.003 \\
    \hline
        Quantum Discord & 0.58 + 0.10 & 0.617  + 0.0023  \\
    \hline
        Steerability Variable & 0.48 + 0.07  & 0.4822 + 0.0015  \\
    \hline
        Conditional Entropy & -0.15 + 0.29 & -0.157 + 0.005  \\
    \hline
        SSRE & 0.38 + 0.15 & 0.388 + 0.003  \\
    \hline
    \end{tabular}
    \caption{Reconstructed quantum information quantities for $\rho_{\kappa}$ with $\kappa=0.2$ with $10^7$ events and 40 pseudo-experiments.} 
    \label{tab:quantumObservables}
\end{table}

\customsection{Conclusion and Outlook}

The flavor space of particles produced in collider environments encodes quantum information as rich as that of the spin space.  Expressing the flavor density matrix for a neutral meson $M$ in terms of the Pauli matrices in the Bloch vector space, we constructed the flavor correlation matrices for the \mm system at a given time $t$.  The crucial observations are that different components in flavor space are related to different physical observables.  Defining the flavor eigenstates along the $z$-direction, the $y$ and $z$ components are measured from the oscillation of the flavor asymmetry, with a sensitivity related to the mass difference $\Delta m$. The $x$ components are measured from the time dependence of the total semi-leptonic decay rate, with a sensitivity related to the decay lifetime difference $\Delta \Gamma$. 

Our analysis provides a systematic way to construct the complete flavor density matrix for an \mm system, and study its production mechanisms, decoherence phenomena, and quantum properties. From the observational point of view, meson systems with suitable $\Delta m/\Gamma$ and $\Delta \Gamma/\Gamma$ are required. We observed that the $B_s^0$ and $K^0_S$ systems are superior to the $B_d^0$ and $D^0$ systems due to their favorable oscillation and decay properties. Through numerical simulations, we demonstrated that the full density matrices can be precisely reconstructed, as shown in Fig.~\ref{fig:fit} and Table~\ref{tab:fit}.  Through the procedure of quantum tomography, one can construct any quantum information variable, such as concurrence, quantum discord, steerability, and the SSRE, as shown in Table~\ref{tab:quantumObservables}. 

A few final remarks are in order. First, as discussed in Refs.~\cite{Genovese:2001pk,Genovese:2001ze,Bramon:2002yg,Genovese:2003dp,Bertlmann:2004cr,Bramon:2004pp,Bramon:2005mg,Genovese:2005vv}, using Bell nonlocality to exclude local hidden variable models is not possible in meson pair systems.\footnote{With an alternative detector setup it may be possible to probe local hidden variable models~\cite{Bramon:2002yg}.}    
Instead, the objective of this work is to reconstruct the complete flavor density matrix of the meson pairs at a given time $t$ at or after they are produced, and study their quantum information properties.
Second, although the overall effects are rather small, the CP-violating interactions in the mixing would change the precession properties in the state evolution, and therefore introduce additional modulation in the observables. The flavor asymmetry observables would involve all of the $3\times 3$ elements of the flavor correlation matrix instead of only the $y$ and $z$ components. Precision tomography could thus be sensitive to new CP-violating sources.  Thirdly, for meson pairs from quark hadronization, there is no first-principles calculation of the flavor state. Our formalism assumes the most general \mm quantum state to begin with as in Eq.~\eqref{eq:rhomesonPair}. This provides a new avenue to explore hadronization mechanisms in terms of quantum correlations, both theoretically and experimentally.

Overall, we demonstrated that performing quantum tomography for the \mm system can facilitate the study of production mechanisms, decoherence phenomena, quantum information variables, and potential new sources of CP violation. 

\begin{acknowledgments}
We thank Alex Lenz, Zoltan Ligeti and Fabio Maltoni for discussions. This work was supported in part by the U.S.~Department of Energy under grant No.~DE-SC0007914 and in part by the Pitt PACC. ML is also supported by the National Science Foundation under grant No.~PHY-2112829.
\end{acknowledgments}

\clearpage
\appendix
\begin{widetext}

\section{Time Evolution}

In this appendix, we list the oscillation parameters and the branching ratios of the relevant mesons in Table~\ref{tab:parameter} and~\ref{tab:BR}, and we summarize the explicit formulas for the meson time-evolution in the following.

\subsection{The Time Evolution of Single Meson} \label{app:evo1}

Consider the density matrix $\rhoM$ for a single meson at time $t=0$.  The time evolution of the density matrix is given by $\rhoM(t)$
\begin{equation}
    \rhoM(t) = \frac{U(t) \rhoM U(t)^\dagger}{\tr (U(t) \rhoM U(t)^\dagger)}.
\end{equation}
The Pauli decomposition of the density matrix parametrizes the quantum state in terms of the Bloch vector $b_i(t)$
\begin{equation}
    \rhoM(t) = \frac{\mathbbold{1}_2 + b_i(t) \sigma_i}{2}.
\end{equation}
Inverting this equation, one can write the Bloch vector as
\begin{equation}
b_i(t)=\tr(\sigma_i\rhoM(t))=\frac{\tr(\sigma_i U(t) \rhoM U(t)^\dagger)}{\tr (U(t) \rhoM U(t)^\dagger)}.
\end{equation}
Using the definition of $U(t)$ from Eq.~\eqref{eq:EvolutionMatrixU} and taking $p=q=1/\sqrt{2}$, the explicit form of the Bloch vector is
\begin{align}
b_x(t) =  \frac{b_x\ch_t -\sh_t }{\ch_t-b_x\sh_t}, \qquad\qquad
b_y(t) =  \frac{b_y\co_t + b_z\si_t}{\ch_t-b_x\sh_t}, \qquad\qquad
b_z(t) =  \frac{b_z\co_t - b_y\si_t }{\ch_t-b_x\sh_t}.
\end{align}
where $\si_t=\sin(\Delta m t)$, $\co_t=\cos(\Delta m t)$, $\sh_t=\sinh(\Delta \Gamma t/2)$, $\ch_t=\cosh(\Delta \Gamma t/2)$, and $b_{i}=b_{i}(0)$ is the component of the Bloch vector at $t=0$.  They are normalized by the fraction of the total number of mesons at time $t$, which is
\begin{equation}
\frac{N(t)}{N_0}= \tr (U(t) \rhoM U(t)^\dagger)= e^{-\Gamma t}(\ch_t-b_x\sh_t).
\end{equation}
Next, we calculate the number distribution of a certain meson state. The ratio of the number of mesons of state $\ket{X}$ at time $t$ over the total number of mesons at $t=0$ is
\begin{equation}
    \frac{N_X(t)}{N_0}=\bra{X}U(t) \rhoM U(t)^\dagger\ket{X} = \tr(\Pi_X \cdot U(t) \rhoM U(t)^\dagger).
\end{equation}
where $\Pi_X= \ket{X}\!\bra{X}$ is the projection operator. For flavor eigenstate $X=M$, $\Pi_M=\Big[\,\begin{matrix}
        1 &  \hspace{-0.3em}0 \\[-1mm]  0 & \hspace{-0.3em}0
    \end{matrix} \,\Big]$, for the other flavor eigenstate $X=\bar{M}$, $\Pi_{\bar{M}}=\Big[\,\begin{matrix}
        0 &  \hspace{-0.3em}0 \\[-1mm]  0 & \hspace{-0.3em}1
    \end{matrix} \,\Big]$.
Therefore, we can construct the flavor asymmetry and total number as
\begin{align}
    \frac{N_M(t)-N_{\bar{M}}(t)}{N_0} &= \tr\Big((\Pi_M-\Pi_{\bar{M}}) \cdot U(t) \rhoM U(t)^\dagger\Big) = \tr(\sigma_z \cdot U(t) \rhoM U(t)^\dagger) \label{eq:NM-NM}, \\
    \frac{N_M(t)+N_{\bar{M}}(t)}{N_0} &= \tr\Big((\Pi_M+\Pi_{\bar{M}}) \cdot U(t) \rhoM U(t)^\dagger\Big) = \tr(\mathbbold{1} \cdot U(t) \rhoM U(t)^\dagger). \label{eq:NM+NM}
\end{align}
The number distributions of the decay products $f$ and $\bar{f}$ can be obtained through the definition of decay rate
\begin{equation}
    \frac{\dd N_{f/\bar{f}}}{\dd t} = \Gamma_{M\to f} N_{M/\bar{M}}(t), \label{eq:decay}
\end{equation}
where we have used the relation that in the CP-conserving case $\Gamma_{M\to f}=\Gamma_{\bar{M}\to\bar{f}}$. Combining Eqs.~\eqref{eq:NM-NM} - \eqref{eq:decay} we find Eqs.~\eqref{eq:trsigmaz} and \eqref{eq:Nf+Nf}.

\begin{table}[h]
    \centering
    \begin{tabular}{|c|c|c|c|c|}
    \hline
     & $B^0_s$  & $B^0_d$ & $D^0$ & $K^0$  \\
    \hline
     $~\Gamma~(\mathrm{ps}^{-1})$  & $0.662$ & $0.658$ & $2.44$ & $5.59\times 10^{-3}$\\
     \hline
     $\Delta m/\Gamma$  & $26.8$ & $0.769$ & $4.6\times10^{-3}$ & $0.95$\\
     \hline
   $\Delta \Gamma /\Gamma$  & $0.135$ & $4.0\times10^{-3}$ & $0.012$ & $1.99$\\
     \hline
    \end{tabular} 
    \caption{Oscillation and decay parameters for neutral mesons \cite{Jubb:2016mvq,HFLAV:2019otj,ParticleDataGroup:2024cfk}.}
    \label{tab:parameter}
\end{table}
\begin{table}[h]
    \centering
    \renewcommand{\arraystretch}{1.4}
    \begin{tabular}{|c|c|c|c|}
        \hline \multicolumn{2}{|c|}{$\mathrm{Br}(M \to f_{\eta_{CP}})$} & \multicolumn{2}{c|}{$\mathrm{Br}(M \to f)$} \\
        \hline $B^0_d \to J/\psi K_S$ & $(8.91\pm0.21)\times 10^{-4}$ & $B^0_d \to \ell^+ \nu_\ell X^-$ & $(20.66\pm0.56)\% $ \\ \hline
        $B^0_s \to J/\psi \eta$ & $(3.9\pm0.7)\times 10^{-4} $ & $B^0_s \to \ell^+ \nu_\ell X^-$ & $(19.2\pm1.6)\% $ \\ \hline
        \makecell{$K_S \to \pi^+\pi^-$ \\ $K_L \to \pi^+\pi^-$} & \makecell{$(69.20\pm0.05)\%$ \\  $(1.967\pm0.010)\times 10^{-3}$ } & \makecell{$K_S \to \pi^\pm \ell^\mp \nu_\ell$ \\ $K_L \to \pi^\pm \ell^\mp \nu_\ell$} & \makecell{$(1.174\pm0.01)\times 10^{-3}$ 
        \\ $(67.59\pm0.13)\% $}\\ \hline
    \end{tabular}
    \caption{$B^0_d$, $B^0_s$, and $K^0$ decay branching fractions used in the analyses~\cite{ParticleDataGroup:2024cfk}.}
    \label{tab:BR}
\end{table}

\subsection{The Time Evolution of a Pair of Mesons}
\label{app:evo2}

Defining the unnormalized density matrix for a pair of mesons at time $(t_1,t_2)$ as 
\begin{equation}
    \RMM(t_1,t_2) \equiv U(t_1) \otimes U(t_2) \,\rhoMM\, U^\dagger(t_1) \otimes U^\dagger(t_2),
\end{equation}
we can write down the meson number distributions similar to Eq.~\eqref{eq:NM-NM} - \eqref{eq:NM+NM}
\begin{align}
    \frac{N_{M\!M}+N_{\bar{M}\!M}+N_{M\!\bar{M}}+N_{\bar{M}\!\bar{M}}}{N_0} &= \tr\Big(\! (\,\mathbbold{1}\;\otimes\;\mathbbold{1}\;) \cdot \RMM(t_1,t_2) \!\Big), \\
    \frac{N_{M\!M}-N_{\bar{M}\!M}-N_{M\!\bar{M}}+N_{\bar{M}\!\bar{M}}}{N_0} &= \tr\Big(\! (\sigma_z\otimes\sigma_z) \cdot \RMM(t_1,t_2) \!\Big), \\
    \frac{N_{M\!M}-N_{\bar{M}\!M}+N_{M\!\bar{M}}-N_{\bar{M}\!\bar{M}}}{N_0} &= \tr\Big(\! (\sigma_z\otimes\;\mathbbold{1}\;) \cdot \RMM(t_1,t_2) \!\Big), \\
    \frac{N_{M\!M}+N_{\bar{M}\!M}-N_{M\!\bar{M}}-N_{\bar{M}\!\bar{M}}}{N_0} &= \tr\Big(\! (\;\mathbbold{1}\;\otimes\sigma_z) \cdot \RMM(t_1,t_2)\! \Big).
\end{align}
To connect to the number distribution of the decay products, we use the definition of decay rate
\begin{equation}
    \frac{\dd^2 N_{ff}}{\dd t_1 \dd t_2} = \Gamma_{M\to f}^2 N_{M\!M}(t_1,t_2).
\end{equation}
Explicitly writing down $\RMM(t_1,t_2)$ using the time evolution operator in Eq.~\eqref{eq:EvolutionMatrixU}, the expansion of $\rhoMM$ in Eq.~\eqref{eq:rhomesonPair}, and calculating the trace, we find the results in Eqs.~\eqref{eq:Ntot} - \eqref{eq:AB}.

\section{Quantum Observables}
\label{app:OQM}

For a two-qubit system, the concurrence $\mathcal{C}$ \cite{Wootters:1997id} is given by 
\begin{equation}
\label{eq:def-concurrence}
\mathcal{C} = \max(0,\lambda_1 - \lambda_2 - \lambda_3 - \lambda_4),
\end{equation}
where $\lambda_i$ ($i=1,2,3,4$) are the eigenvalues, sorted by decreasing magnitude, of the matrix 
\begin{equation}
R_\rho = \sqrt{\sqrt{\rho} \tilde{\rho} \sqrt{\rho}}, 
\qquad\qquad\qquad 
\tilde{\rho} = (\sigma_2 \otimes \sigma_2) \rho^* (\sigma_2 \otimes \sigma_2).
\end{equation}
$0 < \mathcal{C} \leq 1$ indicates an entangled state while $\mathcal{C}=0$ indicates a separable state.  For a separable state, the two subsystems can be fully described independently, but this is not possible for an entangled state.

Bell's inequality is obeyed by all classical theories, including those with local hidden variables~\cite{Bell:1964kc}.  For two qubits, the inequality is given by the Clauser Horne Shimony Holt (CHSH) inequality~\cite{Clauser:1969ny}.  A state that obeys the CHSH inequality is called Bell local while a state that does not is called Bell nonlocal.  The linear approximation to the inequality is measured by the Bell variable $\mathcal{B}$~\cite{Severi:2021cnj}
\begin{equation}
\mathcal{B} = |C_{yy} + C_{zz}| - \sqrt{2}.
\end{equation}
With this normalization, $-\sqrt{2} < \mathcal{B} \leq 0$ indicates a Bell local state and $0 < \mathcal{B} < 2 - \sqrt{2}$ indicates a Bell nonlocal state.

Quantum discord $\mathcal{D}$ is the difference between the total mutual information and the classical mutual information.  The latter requires an extremization over all possible directions, however, for states where only the correlation matrix is non-zero, there is a closed form~\cite{Luo:2008ecu,Han:2024ugl}
\begin{align}
\mathcal{D} =\ &1 + \frac{1}{4}(1 - C_{zz} - C_{xx} - C_{yy}) \log_2 \left( \frac{1 - C_{zz} - C_{xx} - C_{yy}}{4} \right) \notag \\
&+ \frac{1}{4}(1 - C_{zz}  + C_{xx} + C_{yy}) \log_2 \left( \frac{1 - C_{zz} + C_{xx} + C_{yy}}{4} \right) \notag \\
&+ \frac{1}{4}(1 + C_{zz} - C_{xx} + C_{yy}) \log_2 \left( \frac{1 + C_{zz} - C_{xx} + C_{yy}}{4} \right) \notag \\
&+ \frac{1}{4}(1 + C_{zz} + C_{xx} - C_{yy}) \log_2 \left( \frac{1 + C_{zz} + C_{xx} - C_{yy}}{4} \right) \notag \\
&- \frac{1}{2}(1 + \lambda) \log_2 \left( \frac{1 + \lambda}{2} \right)
- \frac{1}{2}(1 - \lambda) \log_2 \left( \frac{1 - \lambda}{2} \right),
\end{align}
where  $\lambda=\mbox{max}\{ |C_{xx}|,|C_{yy}|,|C_{zz}|\}$.  A discord of $\mathcal{D}=0$ indicates a classical state while a discord of $0 < \mathcal{D} \leq 1$ indicates the presence of quantum correlations.

Bell nonlocality is a relatively strict condition that forbids states with local hidden variable descriptions.  A weaker condition is called steerability which forbids states with a description where one subsystem can be described by a local hidden variable theory and the other subsystem has a quantum mechanical description.

The steerability variable $\mathcal{S}$ is derived from the steerability condition~\cite{Jevtic:15}, and is given by
\begin{equation}
\mathcal{S} =\frac{1}{2\pi^2}\int\dd \hat{\mathbf{n}}\sqrt{\hat{\mathbf{n}}^{\mathrm{T}}\mathbf{C}^{\mathrm{T}}\mathbf{C}\hat{\mathbf{n}}} - \frac{1}{\pi},
\end{equation}
where $\mathbf{C}$ is the correlation matrix.  This formulation applies when only the correlation matrix is non-zero.  When $\mathcal{S} \geq 0$ the state is steerable. 

The conditional entropy of a state $\rho_{AB}$ is~\cite{articleCE,Han:2024ugl}
\begin{equation}
S(\rho_{\mathcal{A}}|\rho_{\mathcal{B}})=S(\rho_{\mathcal{AB}}) - S(\rho_{\mathcal{B}}),
\end{equation}
where $S(\rho_{\mathcal{AB}})$ is the Von Neumann entropy of the total quantum state $\rho_{\mathcal{AB}}$ and $\rho_{\mathcal{B}}$ is the reduced density matrix obtained by performing the partial trace over the subsystem $\mathcal{A}$.  The Von Neumann entropy is
\begin{equation}
    S(\rho)=-\tr (\rho\log_2\rho).
\end{equation}
The conditional entropy is the number of bits that would need to be given to one subsystem to reconstruct the other subsystem.  Classically, this ranges from $0$, meaning one subsystem does not need any information to reconstruct the total system, to $1$, meaning that the full information of the other subsystem is required.  Quantum mechanically, the conditional entropy can be as small as $-1$ with the interpretation that a negative value indicates that a bit is available for future quantum communication between the subsystems~\cite{Horodecki:2005vvo}.

In quantum computation, a state that has stronger quantum correlations, like a Bell state, does not necessarily lead to a genuine computational advantage over a classical computer.  States that offer a true advantages are said to possess magic.  A measure of the magic of a quantum state is the second stabilizer R\'enyi entropy (SSRE), is given by~\cite{Leone:2021rzd}
\begin{equation}
    \mathcal{M}_2 = - \log_2 \left( \frac{1+\sum_{i}(b_{i}^{\mathcal{A}})^4 +\sum_{j}(b_{j}^{\mathcal{A}})^4 +\sum_{i,j}C_{ij}^4}{1+\sum_{i}(b_{i}^{\mathcal{A}})^2 +\sum_{j}(b_{j}^{\mathcal{A}})^2 +\sum_{i,j}C_{ij}^2} \right).
\end{equation}
When $\mathcal{M}_2 = 0$ this indicates a state does not possess magic and does not have a computational advantage over a classical system.  The larger the value of $\mathcal{M}_2$ of a state, the larger the computational advantage of the state has.

\begin{figure} 
  \centering
  \includegraphics[width=0.4\linewidth]{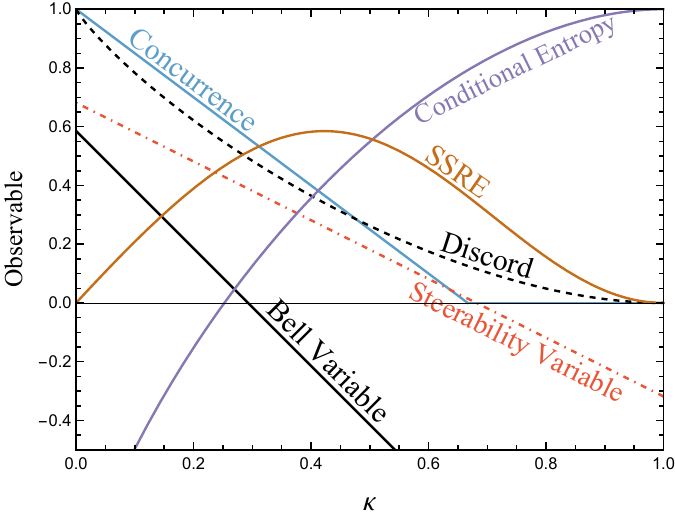}
  \caption{Representative quantum information quantities versus the parameter $\kappa$ for Werner states. }
\label{fig:qm}
\end{figure}

\end{widetext}

\bibliographystyle{apsrev4-1}
\bibliography{ref}
\end{document}